\documentclass[pra,aps,showpacs,twocolumn]{revtex4}

\usepackage{graphicx}
\usepackage{amsmath}
\usepackage{dcolumn}
\usepackage{bm}
\newcommand{\beq}{\begin{equation}}
\newcommand{\eeq}{\end{equation}}
\newcommand{\beqa}{\begin{eqnarray}}
\newcommand{\eeqa}{\end{eqnarray}}
\newcommand{\lam}{\lambda}

\newcommand{\rh}{\rho}
\newcommand{\ga}{\gamma}

\newcommand{\da}{\dagger}

\newcommand{\si}{\sigma}

\newcommand{\om}{\omega}

\newcommand{\la}{\langle}
\newcommand{\ra}{\rangle}

\begin{document}

\date{24th July 2003}

\title{ Qubit Disentanglement and Decoherence via Dephasing}

\author{T.\ Yu$^{1,3}$}
\email{ting@pas.rochester.edu}
\author{J.\ H.\ Eberly$^{1,2}$}
\email{eberly@pas.rochester.edu}
    \affiliation{$^1$Rochester Theory Center for Optical
Science and
Engineering\\
and the Department of Physics \& Astronomy\\
University of Rochester, Rochester, New York 14627\\
$^2$Center for Quantum Information, University of Rochester,
Rochester, New York 14627\\
$^3$Department of Physics, Queen Mary College,
\\University of London, Mile End Road, London E1 4NS, UK }


\begin{abstract}

We consider whether quantum coherence in the form of mutual entanglement 
between a pair of qubits is susceptible to decay that may be more 
rapid than the decay of the  coherence of either qubit individually. 
An instance of potential importance for solid state quantum
computing arises if embedded qubits (spins, quantum dots, Cooper pair 
boxes, etc.)  are exposed to global and local noise at the same time. 
Here we allow separate phase-noisy channels to affect local 
and non-local measures of system coherence. We find that the time for decay 
of the qubit entanglement can be significantly shorter than the time 
for local dephasing of the individual qubits.
\end{abstract}

\pacs{03.65.Ta, 03.65.Yz, 03. 67. -a}
\maketitle  


\newpage

\section{Introduction}
Quantum coherence is fragile in the sense that when a quantum
system is in contact with an environment with many degrees of
freedom, a quantum superposition of a pointer basis inevitably
undergoes a decoherence process \cite{defdeco,zeh}, which is
characterized by the decay of the off-diagonal
elements of the density matrix in the given basis. The fragility
of quantum coherence is responsible for the lack of observation of
Schr\"odinger Cat states in everyday life \cite{str}. Quantum
entangled
states, as special forms of coherent superpositions of multi-partite quantum
states, have recently been recognized as a valuable resource that
is of crucial importance in realization of many quantum
information protocols, such as quantum cryptography
\cite{bbb1,ekert, gisin}, quantum teleportation \cite{bbb2},
superdense coding \cite{bbb3}, and quantum computation
\cite{deu,shor,preskill,mc}. Applications of interest, such as
issues in quantum information processing (QIP)
\cite{de1,kie}, have triggered extensive research aimed at
controlling quantum disentanglement
\cite{shor1,ste,pres,ps,zr,dg,li,pk,viola,viola1}. Apart from the
important link to QIP realizations, a
deeper understanding of entanglement decoherence is also expected
to lead to new insights into the foundations of quantum mechanics
\cite{qub,gisin1}.

The distinctions between ``local" and ``non-local" decoherence can be
important in any application in which coupled qubits may be remote
from one another but nevertheless are coupled or jointly controlled
through common external influences such as electromagnetic fields
(microwave, optical, magnetic, etc.). In this paper, we focus on
these distinctions in the simplest case, a pair of qubits (e.g., spins, 
quantum dots, Cooper pair boxes, etc., which we will refer to as spins) 
embedded in in a solid state matrix and subjected to the most
elementary relaxation mechanism, pure dephasing. However, we include
more than one source of dephasing noise and allow the different
noise sources to produce dephasings with different time scales, in
order to mimic a relatively general practical situation. With
these simple elements we address a few general questions: How does
dephasing affect entanglement as well as local coherence? More
concretely, how rapidly does disentanglement occur compared to the
off-diagonal decay rates of a single qubit density matrix? and how is
the disentanglement rate related to those other coherence decay rates?

Our results, obtained relatively directly through explicit derivation
of the 12 relevant Kraus operators for the several decoherence
channels, are as follows: We show that an environment that causes
pure dephasing can affect entanglement and coherence in very
different ways. As a consequence, we show that disentanglement
generally occurs faster than the decay of the off-diagonal dynamics of
either the composite two-qubit system or an individual qubit. One
manifestation of the difference is that decoherence of a composite
two-qubit state can be incomplete, but entanglement will be
completely destroyed
after a characteristic time. Our findings, while not entirely in
contradiction to one's general intuition, may still come as a
surprise.

Specifically, we will investigate non-local and local aspects of
coherence decay when two qubits are subject to different dephasing
environments. This situation may naturally arise in a long distance
quantum communication, local manipulation of qubits, or
in a local quantum measurement \cite{gisin1,local,vm,vp}. With
application of the quantum map approach to our two-qubit model, we
explicitly show that entanglement, measured by Wootter's
concurrence \cite{woo}, generally decays faster. We show that, by
altering environmental parameters,
the time for disentanglement can be
significantly decreased compared to
the usual time for complete
coherence decay.
The present work extends an earlier result where two qubits are
assumed to interact collectively with a single reservoir \cite{ye}.

The paper is organized as follows. We introduce a two-qubit
dephasing model and describe it in the language of noisy quantum
channels in Sec.~\ref{2qubitsystem} and Sec.~\ref{channel}, respectively.
The explicit solutions to the model are given 
in  Sec.~\ref{explicitsolutions}. In Sec.~\ref{deco}, we study the
dephasing processes for the channel that acts on both qubits and
for the channel that acts on only one qubit. We investigate
disentanglement in Sec.~\ref{entanle}. In Sec.~\ref{fidelity1}, we
discuss coherence decay in terms of transmission fidelity. We comment
and conclude in Sec.~\ref{conc}.

\section{Two-qubit system with three reservoirs}
\label{2qubitsystem}
We consider two qubits $A$ and $B$ that are coupled to a noisy
environment both singly and collectively. One specific realization
would be a pair of spin-$\frac{1}{2}$ particles
embedded in a solid-state matrix and subject to random
Zeeman splittings, for example from random extrinsic magnetic fields.
Analogously, one may envision a pair of polarized photons travelling
along partially overlapping fiber-optic links which are gradually
de-polarized due to random birefringence. In the first case the
relaxation develops in time and in the second case in space. The bare
essentials of the physics can be described by the following
interaction Hamiltonian (which takes $\hbar=1$ and adopts spin
notation):

\begin{equation} \label{hamiltonian}H(t)=-\frac{1}{2}\mu\Big(B(t)(\si_z^A
+\si_z^B)+b_A(t)\si_z^A+b_B(t)\si_z^B\Big),
\end{equation}
where $\mu$ is
the gyromagnetic ratio and $B(t), b_A(t), b_B(t)$ are stochastic
environmental fluctuations in the imposed Zeeman magnetic fields.
Here $\si^{A,B}_z$ are the Pauli matrices:

\begin{equation}
\si^{A,B}_z =
\begin{pmatrix}
1 \,\,& \,\,\,0 \\
0 \,\,& \,\,\,\, -1
\end{pmatrix},\end{equation}
and we take as our standard 2-qubit basis:
\begin{eqnarray}\label{basis}
|1\ra_{AB}&=&|++\ra_{AB},\,
|2\ra_{AB}=|+-\ra_{AB},\nonumber\\
|3\ra_{AB}&=&|-+\ra_{AB},\,
    |4\ra_{AB}=|--\ra_{AB},\end{eqnarray} where $|\pm\pm\ra_{AB}\equiv
|\pm\ra_A\otimes |\pm\ra_B$
    denote the eigenstates of the product Pauli spin
operator $\sigma_z^A\otimes\si_z^B$ with eigenvalues $\pm 1$.

\begin{figure}
\includegraphics{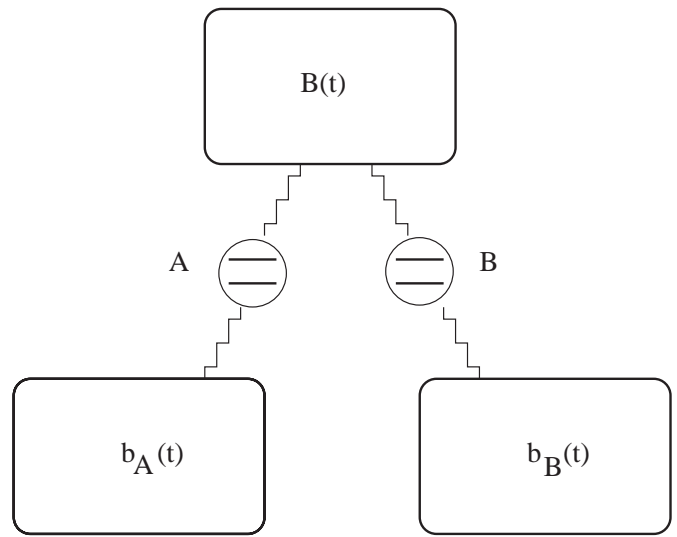}
\caption{\label{fig:epsart} \footnotesize The model consists of
two qubits A and B collectively interacting with the stochastic magnetic
field $B(t)$ and separately interacting with the stochastic
magnetic fields $b_A(t)$ and $b_B(t)$. }
\end{figure}

For simplicity, we assume that the stochastic fields $B(t),
b_A(t), b_B(t)$ are classical and can be characterized as
statistically independent Markov processes satisfying

\begin{eqnarray} <B(t)>&=&0,\\
<B(t)B(t')>&=&\frac{\Gamma}{\mu^2}\delta(t-t'),\label{cor1}\\
<b_i(t)>&=&0, \\
< b_i(t)b_i(t')>&=&\frac{\Gamma_i}{\mu^2}\delta(t-t'),\,\,
i=A,B\label{cor2}
\end{eqnarray}
where $<...>$ stands for ensemble average.
Here $\Gamma$ is the damping rate due to the collective interaction
with $B(t)$, and $\Gamma_{i}$ are the damping rates of qubit $i$ $(i=
A, B)$ due to the coupling to the fluctuating magnetic field
$b_i(t)$. The white-noise properties in (\ref{cor1}) and
(\ref{cor2}) ensure that the two-qubit system will undergo a
Markov evolution. More realistic and more general models are easy
to imagine,  but for the sake of simplicity we will restrict
ourselves to this simple situation.

Hamiltonian (\ref{hamiltonian})
represents a class of models of open
quantum systems that describe a pure dephasing process, which
can arise from interaction with Bosonic environments, independent of
whether they are associated with quantized thermal radiation fields or
phonon fields in solids or other specific physical contexts.

The dynamics under the Hamiltonian (\ref{hamiltonian}) can be
described by a master equation, but for our purpose we find it
more convenient to use the Kraus operator-sum representation
(\ref{sum}). The advantage of the operator-sum representation will
become clear in Sec.~\ref{entanle}. The density matrix for the two
qubits can be obtained by taking ensemble averages over the three noise
fields $B(t), b_A(t), b_B(t)$:

\begin{equation} \rh(t)=<<<\rh_{st}(t)>>>,\label{solution}
\end{equation}
where the statistical density operator $\rh_{st}(t)$ is given by

\begin{equation} \rh_{st}(t)=U(t)\rh(0)U^\dag(t),
\end{equation}
with the unitary operator

\begin{equation} U(t)=\exp\left[-i\int^t_0 dt'H(t')\right].
\end{equation}

\section{Kraus representation for noisy channels}
\label{channel}

We can describe a decoherence process in the language of quantum
channels \cite{wk}. We take a quantum channel to be a
completely positive linear map that acts on the quantum state
space of a system of interest. Let ${\mathcal E}$ be a quantum
channel that maps the input state $\rho_{\rm in}=\rh(0)$ into the
output state $\rho_{\rm out}=\rh(t)$. It is known that the action
of ${\mathcal E}$ can be characterized by a (not unique) set of operators
called Kraus operators $K_\mu$ \cite{kra,preskill,mc}. For any
initial state, the action of quantum map ${\mathcal E}$ is given
by \begin{equation} \label{sum}\rho(t)={\mathcal
E}(\rho)=\sum_{\mu=1}^NK_{\mu}^\dag(t) \rh(0)K_\mu(t), \end{equation} where
$K_\mu$ are Kraus operators satisfying \begin{equation} \sum_\mu K^\dag_\mu
K_\mu=I, \end{equation} for all $t$. Eq.~(\ref{sum}) is often termed the
Kraus (or operator-sum) representation in the literature (for
example, see \cite{preskill,mc}). The operators $K_\mu$ contain
all the information about the system's dynamics.  Unitary
evolution is a special case with only one Kraus operator;
otherwise the channel describes a non-unitary process associated
with damping and decoherence. It can be easily seen that any
$K_\mu$ projects pure states into pure states, but the collective
action of different $K_\mu$'s will typically transform a pure
state into a mixed state.

The most general solution (\ref{solution}) can be
expressed in terms of twelve Kraus operators (under the assumption
that the initial density matrix is not correlated with any of the
three environments):
\begin{equation} \rh(t)={\tilde{\mathcal
E}}(\rh(0))=\sum^2_{i,j=1}\sum^3_{k=1}D_k^\dag E^\dag_j
F^\dag_i\rh(0)F_i E_j D_k, \end{equation}
where the Kraus operators describing the interaction with the
local environmental magnetic fields $b_A(t), b_B(t)$ are given by
\begin{eqnarray}
    \label{k1}E_1
=\left(\begin{array}{clcr}
1 && 0\\
0 && \gamma_A(t)\\
\end{array}
     \right)\otimes I,\, E_2=\left(
\begin{array}{clcr}
0 \,\,&  \,\,\,\, 0 \\

0 \,\,& \,\,\,\om_A(t)\\
\end{array}
     \right)\otimes I,\,\,&&\\
F_1=I\otimes \left(
\begin{array}{clcr}
1 \,\,&\,\,\, 0\\
0 \,\,&\,\,\, \ga_B(t)\\
\end{array}
     \right),\, F_2=I\otimes \left(
\begin{array}{clcr}
0 \,\,&  \,\,\, 0 \\
0 \,\,& \,\,\,\om_B(t)\\
\end{array}
     \right),\,\,&&\label{k5}
     \end{eqnarray}
and the Kraus operators that describe the collective interaction with
the environmental magnetic field $B(t)$ are
given by
    \begin{equation}\label{model1} D_1=\begin{pmatrix}
\ga(t) & 0 & 0 & 0\\
0& 1 &  0 & 0\\
0 & 0& 1& 0\\
0 & 0& 0 & \ga(t)
\end{pmatrix},
\end{equation} \begin{equation}\label{model2} D_2=\begin{pmatrix}
\om_1(t) & 0 & 0 & 0\\
0& 0 &  0 & 0\\
0 & 0& 0& 0\\
0 & 0& 0 & \om_2(t)
\end{pmatrix},
\end{equation} \begin{equation}\label{model3} D_3=\begin{pmatrix}
0 & 0 & 0 & 0\\
0& 0 &  0 & 0\\
0 & 0& 0& 0\\
0 & 0& 0 & \om_3(t)
\end{pmatrix},
\end{equation} Note that the parameters
appearing in
(\ref{k1})--(\ref{model3}) are given by
\begin{eqnarray}
\ga_A(t)&=&{e}^{-{t}/{2T^A_2}},
\,\,\,\ga_B(t)={e}^{-{t}/{2T^B_2}},\label{gaab}\\ 
\om_A(t)&=&\sqrt{1-e^{-{t}/{T^A_2}}},\,\,
\om_B(t)=\sqrt{1-e^{-{t}/{T^B_2}}},\\
\ga(t)&=&e^{-t/2T_2},\,\, \om_1(t)=\sqrt{1-e^{-t/T_2}},\label{para1}\\
\om_2(t)&=&-e^{-t/T_2}\sqrt{1-e^{-t/T_2}},\\
\om_3(t)&=&\sqrt{(1-e^{-t/T_2})(1-e^{-2t/T_2})}, \end{eqnarray} where
$T_2=1/\Gamma$ is the phase relaxation time due to the collective
interaction with $B(t)$ and $T^A_2=1/\Gamma_A$ and
$T^B_2=1/\Gamma_B$ are the phase relaxation times for qubit A and
qubit B due to the interaction with their own environments
$b_A(t), b_B(t)$, respectively. We will no longer write the
time-dependent arguments of the $\ga$'s and $\om$'s as a
notational simplification.

\section{Explicit Solutions, Various Density Matrices}
\label{explicitsolutions}
Three cases can be distinguished:  (1) The case that the two qubits
only interact with their local environments, namely $B(t)=0$; and (2)
the case
in which only one qubit is affected by a local environment, that
is, $B(t)=0$ and one of $b_A(t), b_B(t)$ is also zero.  We call
the above two cases the two-qubits local dephasing channel and the
one-qubit local dephasing channel, respectively. (3) The case that
the two qubits only collectively interact with $B(t)$. We call
this case the collective dephasing channel. We will present these
three cases separately in what follows.

\subsection{Two-qubit local dephasing channel ${\mathcal E}_{AB}$}

When the pair of qubits only interact with the local magnetic
fields, i.e. $B(t)=0$, then the phase-noisy channel, denoted by
${\mathcal E}_{AB}$, is characterized by the following four
composite Kraus operators:
\begin{eqnarray}
    \label{lo1} M_1&=&E_1F_1,\,\,\, M _2=E_1F_2,\,\,\,\\
    \label{lo2}M_3&=& E_2F_1,\,\,\,M_4=E_2F_2,
     \end{eqnarray}
where the fundamental Kraus operators $E_1, E_2, F_1, F_2$ are
defined in (\ref{k1}), and (\ref{k5}). Therefore,
     the effect of
    the quantum channel ${\mathcal E}_{AB}$ on an initial state $\rh$
can be described as follows:
\begin{equation}\label{ch0} \rho(t)={\mathcal
E}_{AB}(\rho(0))=\sum_{\mu=1}^4M_\mu^\da\rho(0)M_\mu. \end{equation}

The explicit solution in the standard basis is given by
\begin{eqnarray}\label{ch} &&\rh(t)={\mathcal E}_{AB}(\rh(0))\nonumber\\
&&=\begin{pmatrix}
\rho_{11} &\ga_B\rho_{12}&  \ga_A\rho_{13} & \ga_A\ga_B\rho_{14}\\
\ga_B\rho_{21} &\rho_{22}&  \ga_A\ga_B\rho_{23}& \ga_A\rho_{24}\\
\ga_A\rho_{31}&\ga_A\ga_B\rho_{32}&\rho_{33}& \ga_B\rho_{34}\\
\ga_A\ga_B\rho_{41}&\ga_A\rho_{42}&\ga_B\rho_{43}& \rho_{44}
\end{pmatrix}.
\end{eqnarray} where $\gamma_A$ and $\gamma_B$ are defined in ({\ref{gaab})
and the notation $\rho_{ij}$ stands for the initial state
$\rh_{ij}(0)$

One immediate conclusion that can be drawn from Eq.~(\ref{ch}) is
that the two-qubit dephasing channel has an effect on all the
matrix elements except for the diagonal elements. Thus,
decoherence takes place in every superposed state of the basis
elements (\ref{basis}). Furthermore, it will turn out later in
Sec.~\ref{entanle} that this channel generally results in faster
disentanglement.

\subsection{One-qubit local dephasing channels  ${\mathcal E}_A$  and
${\mathcal E}_B$}

A simple case arises if the channel is assumed to affect only one
qubit, say qubit A. In this situation,
we call the channel a one-qubit local dephasing channel, and
denote it by ${\mathcal E}_A$.  The one-qubit dephasing channel
${\mathcal E}_A$ can be described by the Kraus operators $E_1,
E_2$. Explicitly, for any initial state $\rh(0)$, the output state
under this channel is given by
\begin{eqnarray} \label{cha}
\rho(t)& =&{\mathcal E}_A(\rho(0))=\sum_{\mu=1}^2E_{\mu}^\dag
\rh(0)E_\mu\nonumber\\
&=&
\begin{pmatrix}
\rho_{11}&\rho_{12}&\ga_A\rho_{13}&\ga_A\rho_{14}\\
\rho_{21}&\rho_{22}&\ga_A\rho_{23}&\ga_A\rho_{24}\\
\ga_A\rho_{31}&\ga_A\rho_{32}&\rho_{33}&\rho_{34}\\
\ga_A\rho_{41}&\ga_A\rho_{42}&\rho_{43}&\rho_{44}
\end{pmatrix}. \end{eqnarray}
Obviously, one can equally well define in terms of Kraus
operators $F_1, F_2$ the one-qubit local dephasing channel ${\mathcal
E}_B$, that has effect only on qubit B.


Since the one-qubit dephasing channels only affect a single
qubit, the coherence of the other qubit is not affected
during processing. Consequently, the coherence  of the
composite two-qubit system cannot be destroyed completely. The
details of the loss of coherence and entanglement under the
channels defined above will be given in Sec.~\ref{deco} and
Sec.~\ref{entanle}.

\subsection{Collective dephasing channel ${\mathcal E}_D$}
If the local stochastic fields $b_A(t), b_B(t)$ are negligible or
switched off, we obtain a model that describes the two qubits
collectively
interacting with a common environment \cite{ye,braun}. In this
case, the solution $\rh(t)$ can be cast into the operator-sum
representation in terms of the three Kraus operators
$D_\mu$: \begin{equation} \label{collective}\rh(t)={\mathcal
E}_D(\rh(0))=\sum_{\mu=1}^3 D^\dag_\mu(t)\rh(0)D_\mu(t), \end{equation}
where $D_\mu,\,\, (\mu=1,2,3)$ are given in
(\ref{model1}),\,\,(\ref{model2}) and (\ref{model3}).

The explicit solution of (\ref{collective}) in the standard basis
(\ref{basis}) can then be expressed as
     \begin{eqnarray}\label{collectivech}
\rho(t)&=&{\mathcal E}_D(\rho(0))\nonumber\\
&=&\begin{pmatrix}
\rho_{11} &\ga\rho_{12}&  \ga\rho_{13} & \ga^4\rho_{14}\\
\ga\rho_{21} &\rho_{22}&  \rho_{23}& \ga\rho_{24}\\
\ga\rho_{31}& \rho_{32}&\rho_{33}& \ga \rho_{34}\\
\ga^4\rho_{41}&\ga\rho_{42}&\ga\rho_{43}& \rho_{44}
\end{pmatrix},
\end{eqnarray} where $\gamma=e^{-t/2T_2}$ is defined in ({\ref{para1}).

As seen from (\ref{collectivech}), the collective dephasing
channel affects the off-diagonal elements $\rh_{14}, \rh_{41}$
more severely than it affects the other off-diagonal elements. As
a result, we have already shown in \cite{ye} that, for a set of
entangled pure states containing $\rh_{14}, \rh_{41}$,
disentanglement proceeds with rates that are faster than the
dephasing rates for qubit A or qubit B. In the case of collective
coupling, the stochastic magnetic field $B(t)$ always affects the
phases of qubit A and qubit B in the same way, so one expects that
for some states the random phases may cancel each other out. Here
such a cancellation due to the underlying symmetry of the interaction 
Hamiltonian can be expected and has been demonstrated in various 
contexts \cite{decohfree}. For the situation considered here, 
the cancellation leads to the existence
of the disentanglement-free subspaces spanned by a basis
consisting of $|2\ra_{AB},\,\, |3\ra_{AB}$. The entangled states
belonging to these subspaces are robust entangled states \cite{ye}.

\section{Local dephasing and mixed dephasing}
\label{deco}

First, we study the coherence decay of a single qubit under the
two-qubit dephasing channel ${\mathcal E}_{AB}$ defined in
(\ref{ch}). The local dephasing rate of the qubit can be determined
from the reduced density matrices $s^A$ and $s^B$,
obtained from the density matrix (\ref{ch}) in the usual way, that
is,
\begin{equation}
s^{A}\equiv{\rm Tr}_B\{\rh\} \quad{\rm and} \quad
s^{B}\equiv{\rm Tr}_A\{\rh\}.
\label{reduced}
\end{equation}
The reduced
density matrices for qubit A and qubit B are thus obtained as:
\begin{equation}
\label{meq1} s^A(t)=
\begin{pmatrix}
\rh_{11}(t)+\rh_{22}(t)\,\, &\,\,    \rho_{13}(t)+\rho_{24}(t)\\
     & \\
\rho_{31}(t)+\rho_{42}(t)\,\,&
\,\,\rho_{33}(t)+\rho_{44}(t)\end{pmatrix},
\end{equation}
and
\begin{equation}
\label{meq2}
s^B(t) = \begin{pmatrix}
\rh_{11}(t)+\rh_{33}(t)\,\, & \,\,   \rho_{12}(t)+\rho_{34}(t) \\
\rho_{21}(t)+\rho_{43}(t)\,\, &\,\, \rho_{22}(t)+\rho_{44}(t)
\end{pmatrix}\quad.
\end{equation}
The dephasing rates are determined by the
off-diagonal elements:
\begin{equation}
s^A_{12}(t)=\rho_{13}(t)+\rho_{24}(t)=\ga_A s^A_{12}(0),
\label{dephas}
\end{equation}
and
\begin{equation}
s^B_{12}(t)=\rho_{12}(t)+\rho_{34}(t)=\ga_B s
^B_{12}(0). \label{dephas1}
\end{equation}
Thus, the dephasing times denoted
by $\tau_{A,B}$ for qubits A and B, respectively,  can
be read off from the expressions for $\ga_A$ and $\ga_B$. That is
\begin{equation}
\label{local}\tau_{A}\equiv \frac{2}{\Gamma_A}, \quad{\rm and} \quad
\tau_{B}\equiv \frac{2}{\Gamma_B}.
\end{equation}

Similar to local dephasing processes, the decoherence rate for the
composite two-qubit system is characterized by the decay rates of
the off-diagonal elements of the density matrix (\ref{ch}). As
seen from (\ref{ch}), the two-qubit off-diagonal elements $\rh_{ij} (i<j)$,
are given by
\begin{equation} \rh_{ij}(t)=e^{-\Gamma_{ij} t} \rh_{ij}(0)
\end{equation}
so the coherence
$\rh_{ij}(t)$ decays on a timescale
\begin{equation} \tau_{\rm dec}=\frac{1}{\Gamma_{ij}},
\end{equation}
where $\Gamma_{12}=\Gamma_{34}=\Gamma_B/2,
\Gamma_{13}=\Gamma_{24}=\Gamma_A/2,
\Gamma_{14}=\Gamma_{23}=(\Gamma_A+\Gamma_B)/2.$

We denote $\tau$ as a time scale on
which all the off-diagonal elements of the density matrix
(\ref{ch}) disappear. The corresponding decoherence rate $1/\tau$ is
called here the mixed dephasing rate in order to distinguish it from
the local dephasing rates of a single qubit. It is evident that the
mixed dephasing rate $1/\tau$ is determined by the slower decaying
elements \cite{com}, and in general, we have,

\begin{equation} \label{chtime}
\tau \geq \tau_A, \tau_B.
\end{equation}
Namely, and this is a key point, the mixed dephasing rate $1/\tau$ is
not shorter than the local dephasing rates for qubit A or qubit B
defined in (\ref{local}). As will be seen in
the next section, this is in contrast with the disentanglement
rate, which can be faster than the local rates $1/\tau_A$
and $1/\tau_B$.

Finally, we study dephasing processes under the local one-qubit
dephasing channels ${\mathcal E}_A$ and ${\mathcal E}_B$. Since
the local dephasing channels only affect one qubit and leave the
other intact, hence some coherence may always exist in the composite
two-qubit state. Let us consider the channel ${\mathcal E}_A$ and a
superposed state of just three members of the standard basis,
\begin{equation}
\label{composite}
|\psi\ra_{AB}=\frac{1}{\sqrt{3}}\Big(|1\ra_{AB}+|2\ra_{AB}+|4\ra_{AB}\Big).
\end{equation}
    Obviously, from (\ref{cha}), we see there is no way that the channel
${\mathcal E}_A$ can destroy the off-diagonal elements $\rh_{12},
\rh_{21}$, so ${\mathcal E}_A$ cannot completely destroy the
coherence of the composite two-qubit state (\ref{composite}).
Next, we look at the dephasing processes for an individual qubit.
Obviously, the coherence of qubit B is not affected by ${\mathcal
E}_A$. That is,
\begin{equation}
s^B(t)=s^B(0).
\end{equation}
Notice that $s^A(t)$
under the channel ${\mathcal E}_A$ is given by (\ref{meq1}).
Hence, as expected, the coherence of qubit A will be destroyed by
the channel ${\mathcal E}_A$ on the dephasing time scale $\tau_A$.
A similar analysis applies to ${\mathcal E}_B$.

\section{Entanglement decay}
\label{entanle}

To describe the temporal evolution of quantum entanglement we have
to measure the amount of entanglement contained in a quantum
state. Since the entanglement decoherence processes are typically
associated with mixed states, we will use Wootter's concurrence to
quantify the degree of entanglement \cite{woo}. The concurrence
varies from $C=0$ for a disentangled state to $C=1$ for a
maximally entangled state. For two qubits, the concurrence may be
calculated explicitly \cite{woo} from the density matrix $\rho$ for
qubits A and B:
\begin{equation}
C(\rh)=\max\left(0,\sqrt{\lam_1}-\sqrt{\lam_2}-\sqrt{\lam_3}
-\sqrt{\lam_4}\,\,\right),
\label{defineconcurrence}
\end{equation}
where the quantities $\lam_i$ are the eigenvalues of the
matrix \begin{equation} \varrho=\rho(\sigma^A_y\otimes
\sigma^B_y)\rho^*(\sigma^A_y\otimes
\sigma^B_y),
\label{concurrence}
\end{equation} arranged in decreasing order.
Here $\rh^*$ denotes the complex conjugation of $\rh$ in the
standard basis (\ref{basis}) and $\si_y$ is the Pauli matrix
expressed in the same basis as:
\begin{equation} \si^{A,B}_y =
\begin{pmatrix}
0 \,\,& \,\,\,-i \\
i \,\,& \,\,\,\, 0
\end{pmatrix}.
\end{equation}
For a pure state $|\psi\ra$, the
concurrence (\ref{defineconcurrence}) is reduced to

\begin{equation}
C(|\psi\ra)=|\la\psi| \si_y^A\otimes
\si_y^B|\psi^*\ra|.\label{concurrence1}
\end{equation}

\subsection{Entanglement decay under two-qubit dephasing
channel ${\mathcal E}_{AB}$}

How does the concurrence behave under a noisy channel? From
(\ref{concurrence1}), for any entangled pure state
\begin{equation} |\Psi\ra=a_1|1\ra_{AB} +a_2|2\ra_{AB}+a_3|3\ra_{AB}
+a_4|4\ra_{AB}\label{generalstate},
\end{equation}
where $\sum_{i=1}^4|a_i|^2=1$,
the concurrence of the pure state (\ref{generalstate}) is given by
\begin{equation}
C(|\Psi\ra)=2|a_1a_4-a_2a_3|.\label{pureconcur}
\end{equation}
It follows from (\ref{ch}) that the concurrence (\ref{pureconcur}) will
always decay to zero on a time scale determined by the dephasing time
$\tau$ defined in (\ref{chtime}). In fact, let us first note that the
density matrix
of (\ref{generalstate}) will approach a diagonal matrix at the
two-qubit dephasing rate $1/\tau$. Since a diagonal density matrix
only describes classical probability, then the concurrence must be
zero.

Moreover, we show in what follows that all entangled (possibly
mixed) states decay at rates that are faster than the
dephasing rates of an individual qubit. For this purpose, we first 
note that the
concurrence $C(\rh)$ is a convex function of $\rho$ \cite{woo};
that is, for any positive numbers $p_\mu$ and density matrices
$\rh_\mu$, $(\mu=1,..,n)$ such that $\sum_\mu p_\mu=1$, one has
\begin{equation} C\Big(\sum_{\mu=1}^n p_\mu
\rh_\mu\Big)\leq \sum_{\mu=1}^n p_\mu C(\rho_\mu).
\end{equation}
Applying the above inequality to (\ref{ch}), we immediately get
\begin{equation}
C(\rh(t)) \leq \sum_{\mu=1}^4C(M_\mu^\dag\rh(0)
M_\mu),\label{ineq}
\end{equation}
where $M_\mu$ are defined in (\ref{lo1}) and (\ref{lo2}).
Now we consider a typical term $C(M_\mu^\dag\rh(0)M_\mu)$ in
(\ref{ineq}) and denote it by
    \begin{equation}
\rh_{\rm out}=M_\mu^\dag \rh(0)M_\mu.\label{out2}\end{equation} Substituting
(\ref{out2}) into (\ref{concurrence}), we have,
\begin{eqnarray} \varrho&=&\rh_{\rm out}(\si^A_y\otimes \si^B_y)\rh_{\rm
out}^*(\si^A_y\otimes\si^B_y)\\ &=& M_\mu^\dag \rho
M_\mu(\sigma^A_y\otimes \sigma^B_y)M^T_\mu
\rho^*M_\mu(\sigma^A_y\otimes \sigma^B_y). \label{con2}\end{eqnarray}
Notice that $\varrho$ has the same eigenvalues as $\varrho'$
\begin{equation} \varrho'= \rho M_\mu(\sigma^A_y\otimes
\sigma^B_y)M_\mu^T \rho^*M^*_\mu(\sigma^A_y\otimes
\sigma^B_y)M_\mu^\dag. \label{con3}\end{equation}
Also note that the Kraus operators defined in (\ref{k1}) satisfy
$M^T_\mu=M_\mu^\dag=M_\mu^*$. Hence it is easy to check that
\begin{equation}
M_\mu(\si^A_y\otimes\si_y^B)M_\mu=\left\{ \begin{array}{ll}
                         \ga_A\ga_B(\si^A_y\otimes\si^B_y) & \mbox{if
$\mu=1$ },\\
                         0 & \mbox{if $\mu\neq 1$}.
                         \end{array}
                         \right.
                         \end{equation}
Then Eq. (\ref{con3}) becomes: \begin{equation}
\varrho''=\left\{\begin{array}{ll}
    \ga^2_A\ga_B^2 \rho(\sigma^A_y\otimes
\sigma^B_y)\rho^*(\sigma^A_y\otimes \sigma^B_y) & \mbox{if
$\mu=1$},\\
0 & \mbox{if $\mu\neq 1$}. \end{array} \right. \label{con4}\end{equation}
By using the definition of concurrence (\ref{concurrence}) and the
inequality (\ref{ineq}), we get: \begin{equation} C(\rh(t))\leq \ga_A\ga_B
C(\rh(0)).\label{final}\end{equation} Hence, the entanglement decay time
$\tau_e$ can be identified as \begin{equation} \label{etime0}
\frac{1}{\tau_e}=\frac{1}{\tau_A}+\frac{1}{\tau_B},\end{equation} where the
local dephasing times $\tau_A, \tau_B$ are given in (\ref{local}). In
this way we immediately see that the
entanglement decay times for the two-qubit channel are shorter
than the local dephasing times $\tau_A, \tau_B$ and hence are
shorter than the mixed dephasing time $\tau$ as well.

By suitably choosing the channel parameters $\Gamma_A, \Gamma_B$,
one can achieve
\begin{equation} \tau_e\ll \tau, \end{equation}
where $\tau$ is the mixed
dephasing time (\ref{chtime}). We have shown here that the decay
time for the entanglement of the two-qubit system can be much
shorter than the mixed dephasing time $\tau$.  An interesting case
occurs when the phasing damping rates for qubit A and qubit B are
assumed the same: $\tau_A=\tau_B$, then from (\ref{etime0}), we
have \begin{equation} \tau_e=\frac{\tau_A}{2}=\frac{\tau_B}{2}=\frac{\tau}{ 2
}.\end{equation} This relation reminds us of the well-known relation between
the phase coherence relaxation rate $T_2 $ and the diagonal
element decay rate $T_1$ in open quantum systems \cite{open}.

To see how the phase-noisy channels affect entanglement in a more
explicit way, it would be instructive to find some examples in
which the concurrence can explicitly be calculated. For this
purpose, let us consider, for example, the following two classes
of almost arbitrary bipartite pure states described by:
\begin{eqnarray}\label{fra2}|\phi_1\ra &=&a_1
|1\ra_{AB} +a_2 |2\ra_{AB}+ a_4 |4\ra_{AB},\\
|\phi_2\ra &=& a_1 |1\ra_{AB} +a_3|3\ra_{AB} +a_4 |3\ra_{AB}
\label{fra3}, \end{eqnarray}
    and
    \begin{eqnarray} \label{fra4}|\psi_1\ra &=& a_1
|1\ra_{AB} +a_2 |2\ra_{AB} + a_3 |3\ra_{AB},\\
|\psi_2\ra&=&a_2 |2\ra_{AB} +a_3|3\ra_{AB} +a_4 |4\ra_{AB}.
\label{fra5} \end{eqnarray} It will turn out in what follows that the
concurrence for those two classes can be calculated exactly.

For the pure state (\ref{fra2}), the concurrence is
$C(|\phi_1\ra)=2|a_1a_4|$. Then the  density matrix with the
initial entangled state (\ref{fra2}) at $t$ is given by \begin{eqnarray}
\label{class1} \rho(t)&=& {\mathcal
E}_{AB}(\rho)\nonumber\\
&=&\begin{pmatrix}
|a_1|^2 & \ga_Ba_1a_2^* & 0 & \ga_A\ga_B a_1a_4^* \\
\ga_B a_2a_1^* & |a_2|^2 & 0 & \ga_A a_2a_4^* \\
0 & 0 & 0 & 0 \\
\ga_A \ga_B a_4a_1^*& \ga_A a_4a_2^* & 0 & |a_4|^2
\end{pmatrix}. \end{eqnarray} The non-zero eigenvalues of the matrix
$\varrho$ defined
in (\ref{concurrence}) are \begin{equation} \lam_1=(1+\ga_A\ga_B)^2|a_1a_4|^2,
\,\,\,\lam_2=(1-\ga_A\ga_B)^2|a_1a_4|^2.\end{equation} Then the concurrence
at $t$ is given by \begin{equation}
C(\rh(t))=2\ga_A\ga_B|a_1a_4|.\label{decay}\end{equation}

Similarly, as another illustration, we may calculate the
entanglement degree of the entangled states (\ref{fra4}). The
density matrix with the initial entangled state (\ref{fra4}) at
$t$ is given by \begin{eqnarray}
\label{class2} \rho(t)&=& {\mathcal E}_{AB}(\rho)\nonumber\\
&=&
\begin{pmatrix}
|a_1|^2  & \ga_B a_1a_2^* & \ga_A a_1a_3^* &0\\
\ga_B a_2a_1^* & |a_2|^2 & \ga_A\ga_B a_2a_3^*&0 \\
\ga_A a_3a_1^* & \ga_A\ga_B a_3a_2^* & |a_3|^2 & 0 \\
    0 &  0 & 0 & 0
\end{pmatrix}.\end{eqnarray}
The concurrence at $t$ can be obtained \begin{equation}
C(\rh(t))=2\ga_A\ga_B|a_2a_3|.\label{decay1}\end{equation} From the above
two examples, we see that the concurrence is determined by the
fast decaying off-diagonal elements $\rh_{14}$ and $\rh_{23}$. The
examples also demonstrate that the bound in (\ref{final}) is the
minimal upper bound.

\subsection{Entanglement decay under one-qubit dephasing channels
${\mathcal E}_A$ or ${\mathcal E}_B$}

Perhaps the
best way of seeing the difference between the disentanglement and
decoherence is to look at the effect of the one-qubit channel on a
two-qubit system. Similar to the two-qubit dephasing channels, it
can be shown that under the one-qubit local dephasing channels,
the concurrence is determined as follows: \begin{eqnarray} C({\mathcal
E}_A(\rh(0))&\leq& \ga_A C(\rh(0)),\\
C({\mathcal E}_B(\rh(0))&\leq& \ga_B C(\rh(0)). \end{eqnarray} Thus, we see
that the one-qubit local dephasing channels can completely destroy
the quantum entanglement after the dephasing times $\tau_A$ or
$\tau_B$ . To be more specific, let us consider the following
entangled pure state: \begin{equation}
|\phi\ra=\frac{1}{\sqrt{3}}\Big(|1\ra_{AB}
+|3\ra_{AB}+|4\ra_{AB}\Big). \end{equation} If we apply the local dephasing
channel ${\mathcal E}_A$ on $\rh=|\phi\ra\la\phi|$ then the output
state is given by \begin{eqnarray} \label{example}
\rho(t)& =&{\mathcal E}_A(\rho)\nonumber\\
&=&
\begin{pmatrix}
\frac{1}{3}& 0 &\frac{\ga_A}{3}&\frac{\ga_A}{3}\\
0 & 0 & 0 & 0\\
\frac{\ga_A}{3} & 0&\frac{1}{3}&\frac{1}{3}\\
\frac{\ga_A}{3}&0 &\frac{1}{3}&\frac{1}{3}
\end{pmatrix}. \end{eqnarray} The concurrence of (\ref{example}) can
be obtained \begin{equation}
C(\rh(t))=\frac{2}{3}e^{-t/2T^A_2}.\end{equation} This means that
disentanglement will proceed exponentially in the local dephasing
time $\tau_A$. As seen from Sec.~\ref{deco}, however, the
coherence of the composite two-qubit state $|\phi\ra$ cannot be
destroyed completely. In general, one-qubit local dephasing
channels ${\mathcal E}_A$ and ${\mathcal E}_B$ cannot destroy the
coherence between different elements of the basis
(\ref{generalstate}). That is \begin{equation} \rh(t) \neq
\sum_{i=1}^4|a_i|^2|i\ra\la i|,\,\,\, {\rm as}\,\, t\rightarrow
\infty.\end{equation} Namely, the decoherence process for the composite
two-qubit system becomes frozen after the local dephasing times
$\tau_A$ or $\tau_B$.

In summary, we have shown that the effect of the dephasing
channels on entanglement and quantum coherence may differ
substantially. In general, the entanglement of two qubits decays
faster than the quantum coherence. In particular, for the channels
${\mathcal E}_A$ and ${\mathcal E}_B$, we have shown that for some
entangled states the local dephasing channels cannot completely
destroy the coherence between the qubits. However, entanglement as
a global property can be completely destroyed by the one-qubit
local dephasing channels.


\section{Input-output fidelity and concurrence}
\label{fidelity1}
It is interesting to look at the coherence and entanglement decays
by comparing input-output fidelity and concurrence. The fidelity
is meant to measure the ``difference'' between an input state and
an output state under a noisy quantum channel. As such, it has
been used to measure the loss of coherence of a quantum state when
the state is subject to the influence of an environment. The
fidelity can be defined as \begin{equation} F(\rh_{\rm in}, \rh_{\rm out})=
\max|\la \psi_1|\psi_2\ra|, \end{equation} where the maximum is taken over
all purifications $|\psi_1\ra,$ $|\psi_2\ra$ of $\rh_{\rm in}$ and
$\rh_{\rm out}$ \cite{jo,shu}. For a pure initial state $\rh_{\rm
in}=|\psi\ra\la \psi|, $ it can simply be written as \begin{equation}
F(\rho_{\rm in},\rh_{\rm out} )=\la\psi |\rho_{\rm out} |\psi\ra.
\label{purefidelity}\end{equation}

In the case of the one-qubit dephasing channel ${\mathcal E}_A$
defined in (\ref{cha}), for the general entangled states
(\ref{generalstate}), Eq.~(\ref{purefidelity}) can be written as
\begin{eqnarray} \label{fidelity} F(\rh, {\mathcal
E}_A(\rh))&=&\sum_{i=1}^4 |a_i|^4\nonumber\\
    &+&2\ga_A \left(|a_1a_4|^2
+|a_2a_3|^2\right)\nonumber\\
&+&2\ga_A\left(|a_1a_3|^2+|a_2a_4|^2\right)\nonumber\\
& +&2 \left(|a_1a_2|^2+|a_3a_4|^2\right).\end{eqnarray} For example, we
consider the entangled pure state (\ref{generalstate}) with
coefficients \begin{equation} a_1=a_2=a_3={\frac{1}{2}},\,\,
a_4=-{\frac{1}{2}}.\end{equation} Then it is easy to check that, for all
times $t$, one has \begin{equation} F(\rh, {\mathcal
E}_A(\rh))\geq\frac{1}{2}.\label{diff1}\end{equation} Hence we again see
that some amount of coherence is always preserved. However, as
shown in the previous discussions, it is obvious that the
disentanglement will be complete; that is, $C$ decays to zero on
the scale of the local dephasing time $\tau_A$: \begin{equation} C(\rh)=1,\,\,
C({\mathcal E}_A(\rh))=0.\label{diff2}\end{equation}

\section{Concluding remarks}
\label{conc}
In this last section we provide some comments on disentanglement,
mixed dephasing processes, and local dephasing processes for an
individual qubit.

It is evident from sections \ref{deco} and \ref{entanle} that in
general entanglement is more fragile than local quantum coherence. In
the case of the examples (\ref{decay}) and (\ref{decay1}), we see
that quantum entanglement is determined by the off-diagonal
elements $\rh_{14}, \rh_{23}$ and their conjugates. In our
two-qubit model, these elements are affected more severely than
other off-diagonal elements. So the disentanglement normally
proceeds in a rate which is faster than the dephasing rate, the
latter being generally determined by the slow-changing off-diagonal
elements $\rh_{12}, \rh_{13}$ etc.

We remark that our simple approach to decoherence in terms of
phase damping channels can easily be extended to other decoherence
channels such as amplitude damping channels. Although the
disentanglement rates are dependent on the choice of the measure
of entanglement, it seems that the fast disentanglement
demonstrated in the paper is of generic character.

Since the local manipulations of a single qubit cannot increase
the entanglement between two qubits, so the quantum map (\ref{ch})
cannot create any entanglement. This is in contrast to the
collective dephasing channel ${\mathcal E}_D$. It is known that
entanglement may be created due to the interaction with a common
heat bath \cite{braun}.

In connection to the foundations of quantum theory, a good
understanding of entanglement decoherence might provide a more
detailed picture of the transition from quantum to classical (or
semiclassical) dynamics. As suggested in this paper, a quantum
system with several constituent particles, if initially in an
entangled state, may undergo several stages of transition,
including the ``collapse'' of entanglement, and then the
``collapse'' of coherence among individual particles. This is a
topic to be addressed in future publications.

Summarizing, our primary interest in this paper is to investigate
the fragility of quantum entanglement. Through simple two-qubit
systems we have shown that the most commonly assumed (dephasing)
model of noisy environments may affect entanglement and quantum
coherence in a very different manner. While preserved in the case
of collective interaction, the permutation symmetry of the two
qubits is violated by local environment interactions. As a result,
there are no entanglement decoherence-free subspaces for the local
dephasing models, which means that entanglement for all entangled
states will decay into zero after some time. Our results suggest
that, when spins are coupled to inhomogeneous external magnetic
fields, disentanglement and dephasing may proceed in rather
different time scales. It means that, even in the situation that
quantum coherence has relatively long relaxation times,
entanglement may be rapidly suppressed. This issue is certainly of
importance in realizing qubits as spins of electrons or nuclei in
solids.


\section*{Acknowledgments}

We acknowledge financial support from the NSF Grant
PHY-9415582, the DoD Multidisciplinary University Research
Initiative (MURI) program administered by the Army Research Office
under Grant DAAD19-99-1-0215, and a cooperative research grant from
the NEC Research Institute. One of us (TY) would also like to thank
the Leverhulme Foundation for support. We are grateful to several
colleagues for questions, comments and discussions: Lajos Diosi, Ian
Percival, Krzysztof Wodkiewicz.

\bibliography{apssamp}

\begin{thebibliography}{10}
\bibitem{com} Obviously, there exist some special entangled states,
such as the Bell states
$|\Psi_{\pm}\ra=\frac{1}{\sqrt{2}}(|1\ra_{AB}\pm |4\ra_{AB})$ and
$|\Phi_{\pm}\ra=\frac{1}{\sqrt{2}}(|2\ra_{AB}\pm|3\ra_{AB})$,
where the dephasing rates are determined by $\ga_A\ga_B$ terms
appearing in (\ref{ch}). Throughout the paper, we will be
interested in more general cases.
\end{thebibliography}

\end{document}